\documentstyle[preprint,aps]{revtex}
\begin{document}
\draft
\title {\Large \bf 
Superfluidity in sympathetic cooling with atomic Bose condensates} 
\author{E. Timmermans and R. C\^{o}t\'{e} \\
Institute for Atomic and Molecular Physics \\
Harvard-Smithsonian Center for Astrophysics \\
60 Garden Street \\
Cambridge, MA 02138} 
\date{\today}
\maketitle
\begin{abstract}

	The dynamical structure of an atomic Bose-Einstein condensate
limits the efficiency of the condensate
in cooling slow impurity atoms.  To illustrate the
point, we show that an impurity atom moving in a homogeneous 
zero-temperature condensate is not scattered incoherently if its velocity
is lower than the condensate sound velocity $c$, limiting cooling to
velocities $v \geq c$.  This striking effect is an expression of
superfluidity and provides a direct means to detect the fundamental
property of superfluidity in atomic condensates.  Furthermore, we
show that the fermionic lithium-isotope, $^{6}$Li,
is a reasonable candidate for sympathetic cooling by a $^{23}$Na-condensate.

\end{abstract}


	The significance of the observation of Bose-Einstein condensation
in alkali-atom traps \cite{BECf} can hardly be overstated : these experiments 
represent the first explorations into the realm of dilute atomic
quantum gases.  It is thus unfortunate that the same technology is not
expected to cool fermionic atoms into the degenerate regime.
The problem stems from the antisymmetry which forbids $s$-wave scattering
of indistinguishable fermions, implying a vanishing thermalization rate
in evaporative cooling at low temperatures \cite{sym1}.  
To overcome this difficulty,
the JILA-collaboration proposed an interesting alternative \cite{sym} : 
the fermions can be cooled by bringing them in contact with 
an atomic
condensate.  This scheme of sympathetic cooling would
be an interesting and important application of BEC, cooling
not only fermionic alkali atoms, but any gas that cannot be
cooled by means of the traditional cooling techniques, such as inertial gas
atoms or perhaps even molecules \cite{Doyle}.

        In this letter, we discuss the efficiency of such cooling schemes
\cite{Lew}.
Whereas one generally expects sympathetic cooling to be most efficient
with a cooling system of lowest possible temperature, we point out that this
is not necessarily true for a Bose condensate.  
In fact, we show that for a zero-temperature
condensate, the cooling of a low-density gas of `impurity atoms'
(distinguishable from the condensate atoms) exhibits a limit due to the
superfluidity of the condensate: impurity atoms slower
than the condensate sound velocity are not scattered incoherently
and cannot be cooled down to lower velocities.
At finite temperatures the condensate contains
a normal fluid fraction and the impurity atoms
cool down further, but at a reduced rate.  
This aspect of sympathetic cooling provides the opportunity to
observe the fundamental property of superfluidity.  
 
	At zero temperature the condensate scatters impurity atoms by creating
quasi-particles.  In a condensate of bosons with mass $m_{B}$ and density
$n_{B}$, the energy dispersion of the quasi-particles is $\omega_{q}
= cq \sqrt{ 1 + (q/k_{c})^{2}}$, where $k_{c}
= \sqrt{16\pi n_{B} a}$ is the
inverse coherence length and
$c = k_{c}/2m_{B}$ the condensate sound velocity.  In the long wavelength limit,
$q \ll k_{c}$, the momentum region of importance in describing slow
impurity atom scattering, $\omega_{q} \simeq cq$ 
is phonon-like.  We can then repeat
Landau's argument for the absence of resistance in superfluid flow,
which translates here into the absence of incoherent scattering for
impurity atoms moving at a velocity $v < c$.
The argument is based
on the observation that energy and momentum cannot be simultaneously conserved
if $v < c$.  To see this \cite{Huang}, 
consider the energy transfer $\Delta E = c 
\sum_{n} |{\bf q_{n}}|$, where the ${\bf q}_{n}$-vectors represent
the momenta of the phonons created in the scattering, giving a total
momentum transfer ${\bf q} = \sum_{n} {\bf q}_{n}$.  From the energy
difference experienced by the scattered impurity atom of mass $m_{I}$, 
$\Delta E =
{\bf v} \cdot {\bf q} - q^{2}/2m_{I} < v  q$ and since $q \leq 
\sum_{n} |{\bf q}_{n}| = \Delta E/c$, we find that $\Delta E < (v/c) \Delta E$,
which cannot be satisfied if $v < c$.

        To obtain a quantitative understanding of BEC sympathetic cooling, we
discuss a somewhat idealized model of a homogeneous 
zero-temperature BEC, cooling an impurity-atom system of such
low density that
1. the interactions between impurity atoms
can be neglected, 2. the impurity system remains non-degenerate
throughout the cooling process and 3. we can neglect the heating of the
condensate.  
The occupation number $n_{\bf k}$ of the plane wave state of momentum
${\bf k}$ evolves in time according to a simple Boltzmann equation :
\begin{equation}
\dot{n}_{\bf k} = \sum_{\bf q}  \{
-w({\bf k};{\bf k}-{\bf q}) \; n_{\bf k}  +
w({\bf k}+{\bf q} ; {\bf k}) \; n_{{\bf k}+{\bf q}}
\}  ,
\label{e:bolt1}
\end{equation}
where $w({\bf k}_{in} ; {\bf k}_{f})$ is the rate for
scattering of an impurity atom from an initial momentum
${\bf k}_{in}$ to momentum ${\bf k}_{f}$.  
The $w$-rate is proportional to the dynamical structure of the scattering
(or cooling) system.  We derive this result for an impurity atom that
interacts weakly with the bosons by means of a pseudo-potential,
$H' = \lambda \sum_{i=1}^{N} \delta ({\bf r} - {\bf r}_{i})$, where
${\bf r}$ represents the impurity atom position and ${\bf r}_{i}$ the position
of the $i$-th boson, and where $\lambda = 2 \pi a_{IB}/\mu$, with $a_{IB}$ the
scattering length for the impurity-boson scattering and $\mu$ the reduced
mass, $\mu^{-1} = m_{B}^{-1} + m_{I}^{-1}$.  The interaction matrix element
between the initial state, $|\Psi_{0},{\bf k}_{in}\rangle 
\equiv |\Psi_{0}\rangle \otimes |{\bf k}_{in}\rangle$, where $|\Psi_{0}\rangle$
denotes the BEC ground-state, and the final state 
$|\Psi_{f},{\bf k}_{f} \rangle \equiv |\Psi_{f}\rangle \otimes |{\bf k}_{f}
\rangle$, is then equal to
\begin{eqnarray}
\langle  \Psi_{f}, {\bf k}_{f}| H' |\Psi_{0},{\bf k}_{in} \rangle
&=& \frac{\lambda}{V} 
\int \! d^{3} r_{1} \ldots d^{3} r_{N}  \Psi_{f}^{*}
({\bf r}_{1}, \ldots ,{\bf r}_{N})
\nonumber \\
&& 
\times  \sum_{j} \exp(i{\bf q}\cdot {\bf r}_{j}) 
\Psi_{0} ({\bf r}_{1}, \ldots ,{\bf r}_{N})
\nonumber \\
&=& \frac{\lambda}{V}
\langle \Psi_{f}|\rho_{\bf q}|\Psi_{0} \rangle \; \; \; ,
\label{e:intm}
\end{eqnarray}
where ${\bf q}$ is the momentum transfer ${\bf k}_{in}-{\bf k}_{f}$,
$\rho_{\bf q}$ the Fourier transform of the particle density 
and V the macroscopic volume of the homogeneous system.
Representing the initial and final state energies by $E_{0} + \epsilon
({\bf k}_{in})$ and $E_{f}+ \epsilon ({\bf k}_{f})$, 
we obtain for the scattering rate in a Fermi-Golden
rule calculation:
\begin{eqnarray}
w({\bf k}_{in} ; {\bf k}_{f}) &=&
2\pi \sum_{|\Psi_{f} \rangle}
| \langle \Psi_{f},{\bf k}_{f} | H' | \Psi_{0},{\bf k}_{in} \rangle
|^{2} \;
\nonumber \\
&& \; \times \; \delta (E_{f}+\epsilon ({\bf k}_{f}) -
[E_{0} + \epsilon ({\bf k}_{in})]) 
\nonumber \\
&=& 2 \pi \left( \frac{\lambda}{V} \right) ^{2} \; S({\bf q},
[ \epsilon ({\bf k}_{in}) - \epsilon ({\bf k}_{f})]) \; ,
\label{e:wrate}
\end{eqnarray}
where we introduced the dynamical structure factor 
$S({\bf q},\omega) = \sum_{|\Psi_{f}\rangle}
|\langle \Psi_{f}|\rho_{\bf q}|\Psi_{0}\rangle |^{2}
\delta (E_{f}-E_{0}-\omega)$. 
In the thermodynamic limit, $\sum_{\bf q} \rightarrow
V/(2\pi)^{3} \int d^{3} q$, 
the Boltzmann-equation (\ref{e:bolt1}) takes on the following form :
\begin{eqnarray}
\dot{n}_{{\bf k}}(t) &=& \left( \frac{\lambda}{2\pi} \right)^{2}
 \left[ 
- \int d^{3} q \; s({\bf q}, {\bf v} \cdot {\bf q} - q^{2}/2m_{I})
 n_{\bf k}(t)
\right.
\nonumber\\
&& \left. + \int d^{3} q \; s({\bf q},{\bf v} \cdot {\bf q} + q^{2}/2m_{I})
\; n_{{\bf k}+{\bf q}}(t)
\right] , 
\label{e:boltc}
\end{eqnarray}
where $s$ denotes the dynamical structure
factor density,  $s({\bf q},\omega) = [S({\bf q},\omega)/V]$. For
a zero-temperature dilute condensate $s$ is approximated accurately by
\cite{Pines}:
\begin{equation}
s({\bf q},\omega) \cong n_{B} \; \frac{q^{2}/2m_{B}}{\omega_{q}} \;
\delta (\omega - \omega_{q}) \; \; , 
\label{e:s}
\end{equation}
describing the creation of a single quasi-particle of momentum ${\bf q}$
and energy $\omega_{q}$ (defined above).
Although the above derivation assumes weak interaction between the impurity
and the condensate (reasonable if $v > c$),
the rigorous derivation of the quantum Boltzmann equation \cite{Lang-Dan} shows
that Eq.({\ref{e:boltc}) has a broader range of validity.

	With Eqs.(\ref{e:boltc}) and (\ref{e:s}), we calculate the rate 
$dQ/dt = - \int d^{3} k \epsilon ({\bf k}) \dot{n}_{\bf k} $ at which energy is 
transferred from the impurity atoms to the condensate,
\begin{equation} 
\frac{d Q}{d t} = \frac{a_{IB}^{2}}{\mu^{2}} \int \! d^{3} k \,
    n_{\bf k} \int \! d^{3}q \, \omega_{q}  s({\bf q},
[\epsilon({\bf k})-\epsilon
    ({\bf k} - {\bf q})]) , \label{e:dq1}
\end{equation}
where $n_{\bf k}$ is normalized so that 
$\int d^{3} k \; n_{\bf k}$ equals the number of impurity atoms.  
Performing the remaining q-integration,
we find
\begin{equation} 
\frac{d Q}{d t} = 4 \pi a_{IB}^{2} n_{B} \frac{2 \mu^{2}}{m_{I} m_{B}} \;
    \int d^{3} k \; n_{\bf k} \frac{k}{m_{I}} \frac{k^{2}}{2m_{I}} F_{C}(k) ,
    \label{e:dq2}
\end{equation}
where} $F_{C}(k)=0$ if $v<c$, and
\begin{equation} F_{C}(k) = \left( \frac{1-\sqrt{1-
    \left[ 1 - \left( \frac{m_{I}}{m_{B}} \right)^{2}\right]
\left[ 1 - \left( \frac{c}{v} \right)^{2} \right] } } 
     { \left[ 1 - \left( \frac{m_{I}}{m_{B}} \right) \right] }
\right) ^{4}
\end{equation}
if $v>c$.  A classical system of noninteracting particles of mass
$m_{B}$ has a dynamical structure factor density
$s({\bf q},\omega) = n_{B} \delta(\omega-q^{2}/2m_{B})$,
giving the above result with $F_{C} \rightarrow 1$.  Thus the $F_{C}(v)$
-function, shown in Fig.1, corrects for the condensate dynamical structure.
At $v \gg c$, $F_{c}(v) \simeq 1$ and 
the energy transfer rate for cooling fast impurity atoms by a condensate
is the same as the rate for cooling by noninteracting atoms at rest.
At $ v \geq c$, the condensate dynamical structure factor decreases
the rate of energy transfer until at $v < c$ the impurity atoms are not
scattered and no transfer of energy takes place, confirming the picture
derived from the Landau argument above.

	Even if interactions between the impurities cannot be
neglected (giving impurity-impurity collision terms in the Boltzmann
equation), the heat transfer rate of Eq.(\ref{e:dq2}) is correct, provided
the impurity system is non-degenerate and interacts weakly with the
condensate.   In the limit of strong
impurity-impurity interactions, 
these interactions `thermalize' the distribution $n_{\bf k}$ 
during the cooling process and $n_{\bf k}(t)$
remains Maxwellian,
$n_{\bf k}(t) \sim \exp \left[ -k^{2}
/2m_{I}k_{B}T(t) \right]$, where $k_{B}$ denotes the Boltzmann constant.
We can then calculate the
rate of temperature decrease in accordance with the heat transfer rate,
$dT/dt \simeq - [dQ/dt]/C$, where $C$ is
the heat capacity, $C = \frac{d}{dT} \int d^{3} k \; n_{\bf k} \epsilon
({\bf k})$ $= \int d^{3} k \;
 \epsilon ^{2} ({\bf k}) n_{\bf k} /k_{B}T^{2}$.  This gives
$dT/dt \simeq - T/\tau_{c}$, where the cooling rate $\tau_{c}^{-1}$
depends on the thermal velocity, $v_{T} = \sqrt{2m_{B}k_{B}T}$,
as $\tau_{c}^{-1} = (\mu^{2}/m_{I}m_{B}) \;
(16/15\sqrt{\pi}) \times 4 \pi n_{B} a_{IB}^{2} v_{T} \; F_{S}(v_{T})$
$ \sim (\mu^{2}/m_{I} m_{B}) \times  4 \pi n_{B} a_{IB}^{2} v_{T}
\; F_{S}(v_{T})$, where the $F_{S}$-factor corrects for 
the condensate structure factor, expressing the corresponding
suppression of the cooling rate, 
$F_{S} = \int d^{3} k \; n_{\bf k} k^{3} F_{C}(k) \; / \int d^{3} k \;
n_{\bf k} k^{3}$ (see Fig.1).

	However, the interaction with a dense
condensate generally modifies the condensate.  We return, therefore, to
to the low-density impurity system where $n_{\bf k}(t)$
does not remain Maxwellian (although the above 
cooling rate is still a reasonable estimate) and we solve the
Boltzmann equation numerically.
In Fig.2, we show the time evolution of an isotropic impurity
distribution that is cooled by a zero-temperature condensate.
The distribution starts out as a Maxwellian with $v_{T} = 3 c$,
and progressively narrows in velocity-space but does not converge to zero 
velocity
-- instead the position of its maximum converges to a value slightly below c:
the signature
of superfluidity.  The distributions are plotted at times $t=0.4 n \tau$,
$n=0,\cdots,10$, where $\tau^{-1}$ is the cooling rate for $v_{T} = c$,
$\tau^{-1} = 4 \pi a_{IB}^{2} n_{B} 
c (\mu^{2}/m_{I} m_{B})$.  The
mass-ratio of the impurity and condensate atoms in the calculation is $6/23$,
corresponding to the fermionic $^{6}$Li-isotope cooled by a
$^{23}$Na-condensate.

	The choice of the elements and their atomic states 
is crucial in sympathetic cooling since processes such as spin-flip scattering 
limit the lifetime of the impurity system in contact with the condensate.
$^{23}$Na, the atom used in the high-density BEC-systems created at MIT,
and $^{6}$Li 
are promising choices:  their atomic properties are well understood 
and the spin-flip rates relatively small.   

 	The magnetic traps in which BEC has been realized confine
only a fraction of the 
hyperfine states, namely the low magnetic field seekers.
At small field intensities, the low field seeker
states are $(f$ $m) = (\frac{3}{2}$ $\frac{3}{2})$,
$(\frac{3}{2}$ $\frac{1}{2})$ and
$(\frac{1}{2}$ -$\frac{1}{2})$ for $^{6}$Li,
and (2 2), (2 1) and (1 -1)
for $^{23}$Na. In the elastic approximation \cite{dalgarno65}, also known as 
the Degenerate Internal States (DIS)
approximation \cite{verhaar87}, one describes the scattering of atoms in these
hyperfine states in terms of the singlet and triplet scattering amplitudes.
At the relevant low energies, the scattering is dominated 
by the $s$-wave contribution and the triplet (singlet)
phase shift $\delta_{T(S)}$ can be represented
by $-ka_{T(S)}$, where $k$ is the relative momentum and $a_{T(S)}$ is
the scattering length for the triplet (singlet) scattering 
(see Table~\ref{table1}).
The scattering length for elastic processes (responsible for
cooling and thermalisation) is then $a\simeq P_{S}a_{S} + P_{T}a_{T}$
where $P_{S}$ and $P_{T}$ represent the probability that the scattering
product is in the singlet and triplet state.
For example, $^{23}$Na-$^{23}$Na scattering in (1 -1)+(1 -1) gives
$a\simeq \frac{13}{16} a_{T}+\frac{3}{16}a_{S}=69 a_{0}$,
in good agreement with $86^{+66}_{-23}a_{0}$ \cite{verhaar}
and close to $52\pm 5 a_{0}$ \cite{tiesinga}.
The spin-flip cross section is 
\begin{equation} \sigma_{sf} = M_{sf} \; \pi (a_{T}-a_{S})^{2} ,\label{sigma-sc} \end{equation}
where $M_{sf}$ is a coefficient that depends on
the specific states involved in the process.
This multiplicative factor vanishes if the initial state of the scatterers is
of a pure triplet or singlet character, implying that it is possible to
choose the atomic states so as to avoid spin-flip processes all together.
In cases where spin-flip occurs, their rate may still be reduced
if the singlet and triplet scattering lengths are nearly equal, as for
the $^{87}$Rb-condensates observed at JILA \cite{sym,Rb-symp,Rb-symp-jila}.

	For $^{6}$Li -$^{6}$Li\ collisions, only collisions between $(f$ $m)$
states
($\frac{3}{2}$ $\frac{3}{2}$)+($\frac{3}{2}$ $\frac{3}{2}$),
($\frac{3}{2}$ $\frac{1}{2}$)+($\frac{3}{2}$ $\frac{1}{2}$), and
($\frac{1}{2}$ -$\frac{1}{2}$)+($\frac{1}{2}$ -$\frac{1}{2}$)
are without spin-flip.
Because of the fermionic nature of $^{6}$Li, these
are all forbidden to $s$-wave collisions (no cooling possible).
For $^{23}$Na-$^{23}$Na collisions, the only scattering products not leading to spin-flip
decay are (2 2)+(2 2) and (1 -1)+(1 -1).
The collision channel (2 2)+(2 2) is purely elastic
with $a$ given by $a_{T}$ in Table~\ref{table1}.
(1 -1)+(1 -1) is a mixture of both singlet and triplet and many
exit channels could lead to trap loss. However, at low temperature,
all these channels are closed.
 
For $^{23}$Na-$^{6}$Li collisions, the combinations of states without
spin-flip are ($\frac{3}{2}$ $\frac{3}{2}$)+(2 2) and
($\frac{1}{2}$ -$\frac{1}{2}$)+(1 -1). Only elastic collisions
are possible for ($\frac{3}{2}$ $\frac{3}{2}$)+(2 2)
but for 
($\frac{1}{2}$ -$\frac{1}{2}$)+(1 -1), many decay channels exist, 
although 
not physically accessible at low temperatures.

The scattering lengths for $^{6}$Li reported in Table~\ref{table1}
were derived by Abraham {\it et al.} \cite{huletli6} and those for
$^{23}$Na were calculated by C\^{o}t\'{e} and Dalgarno \cite{cote95}.
They are in agreement with recent estimates \cite{verhaar,tiesinga}.
For the mixed case $^{23}$Na-$^{6}$Li, we
computed the scattering lengths using singlet and triplet model potentials
formed by the extension of {\it ab initio} data with long-range
tails. The {\it ab initio} data originate from
Schmidt-Mink {\it et al.} \cite{schmidt-mink} 
and extend to large distances
(30$a_{0}$). Beyond the {\it ab initio} region, the potential
is described by the
dispersion coefficients of Marinescu {\it et al.} \cite{mircea} and takes 
the form
$-C_{6}/r^{6}-C_{8}/r^{8}-C_{10}/r^{10} \mp Ar^{\alpha}e^{-\beta r}$.
The $\mp$ sign of the exchange term stands for the singlet and 
triplet potential respectively. The parameters were
computed using expressions given
by Smirnov and Chibisov \cite{smirnov} and take on the 
following values (in atomic units):
$A=0.0124$, $\alpha = 4.626$ and $\beta =1.2445$.
The two regions were smoothly joined using a cubic spline fit.

The spin-flip rate coefficients (if spin-flip scattering occurs 
and $M_{sf} \sim 1$) can be
estimated with the values in Table~\ref{table1} by
${\cal R} = \langle v \sigma_{sc} \rangle $ where $v$ is the velocity
of the pair of atoms undergoing a spin-flip,
and where the average is taken over a Maxwellian velocity distribution.
For $^{6}$Li-$^{6}$Li, ${\cal R}$ is
quite large due to its large scattering length,
and still sizeable for $^{23}$Na-$^{23}$Na, but one order of magnitude smaller
for $^{23}$Na-$^{6}$Li (see Table~\ref{table-R}).
This situation is similar to the case of $^{87}$Rb, where
${\cal R}\sim 2.2\times 10^{-14}\mbox{cm}^{3}/s$ and sympathetic
cooling of two different hyperfine states has been observed \cite{sym},
presumably because the
singlet and triplet scattering lengths are nearly equal.
This implies that the four
other possible collision channels for $^{23}$Na-$^{6}$Li, 
($\frac{3}{2}$ $\frac{1}{2}$)+(2 2), ($\frac{1}{2}$ -$\frac{1}{2}$)+(2 2),
($\frac{3}{2}$ $\frac{3}{2}$)+(1 -1) and ($\frac{3}{2}$ $\frac{1}{2}$)+(1 -1),
could lead to sympathetic cooling, as long as the  
$^{6}$Li states are not mixed (otherwise
spin-flip among $^{6}$Li states will take place).
In that case, sympathetic cooling is feasible if the cooling rate
$\tau_{c}^{-1}$ exceeds the rate of spin-flip scattering $n_{B} {\cal R}$ or
\begin{equation}
M_{sf} \left( \frac{a_{IB}^{T} - a_{IB}^{S}}{a_{IB}} \right)^{2} 
\left( \frac{ \mu^{2} }{m_{I} m_{B} } \right) \; \ll \; 1 \; ,
\label{e:feas}
\end{equation}
where the $_{IB}$ subsripts denote that the scattering lengths are calculated
for impurity-boson scattering.

Finally, we remark that the best option for sympathetic cooling of $^{6}$Li by 
$^{23}$Na in the (1 -1) state (such as in the MIT condensate) 
is to trap $^{6}$Li in the state
($\frac{1}{2}$ -$\frac{1}{2}$); this 
ensures a small trap loss from spin-flip. On the
other hand, a condensate of $^{23}$Na--atoms in the (2 2)-state with
$^{6}$Li in the ($\frac{3}{2}$ $\frac{3}{2}$)-state is the best option since
no spin-flip is possible.
In both cases, there is a sizeable scattering length, $a_{IB}$, hence  
efficient sympathetic cooling.

	In conclusion, we have shown that the superfluidity of atomic Bose
condensates limits their efficiency for sympathetic cooling. This effect 
provides a possibility for direct observation of the superfluidity of
the dilute atomic trap condensate systems.  Furthermore, we have discussed
the feasibility of such sympathetic cooling, and its dependence on the
atomic scattering parameters.  For the special case of
cooling $^{6}$Li and $^{23}$Na, we showed 
that sympathetic cooling is possible and we
discussed the best options for the choice of hyperfine states to be trapped.
 
        The work of both authors was supported by the
NSF through a grant for the
Institute for Atomic and Molecular Physics at Harvard University
and Smithsonian Astrophysical Observatory.

\newpage

\begin{table}
\protect\narrowtext
\begin{tabular}{cccc}
      & $^{6}$Li & $^{23}$Na  & $^{23}$Na-$^{6}$Li \\ \hline & & & \\
$a_{T}$   & $-2160$  & 77.3 & 31.1 \\
$a_{S}$   & 45.5  & 34.9    & 39.2 \\
& & & \\
\end{tabular}
\caption{ \protect\narrowtext
        Singlet and triplet scattering lengths in 
        Bohr radii $a_{0}$,
        for isotopically pure and mixed alkali gases.
        }
\label{table1}
\end{table}

\begin{table}
\protect\narrowtext
\begin{tabular}{cccc}
 $T ($K$)$  & $^{6}$Li-$^{6}$Li & $^{23}$Na-$^{23}$Na  & $^{23}$Na$-$$^{6}$Li   \\ \hline
  & & \\
 $10^{-7}$  & $1.1\times 10^{-9}$ & $2.1\times 10^{-13}$ & $1.2\times 10^{-14}$
\\
 $10^{-6}$  & $3.6\times 10^{-9}$ & $6.8\times 10^{-13}$ & $3.8\times 10^{-14}$
\\
 $10^{-5}$  & $1.1\times 10^{-8}$ & $2.1\times 10^{-12}$ & $1.2\times 10^{-13}$
\\
 & & & \\
\end{tabular}
\caption{ \protect\narrowtext
          Spin-flip rate coefficients ${\cal R}$ for collisions between
          $^{6}$Li and  $^{23}$Na atoms in units of cm$^{3}/$s.
        }
\label{table-R}
\end{table}

\newpage
\centerline{\bf Figure Captions}
\noindent
\underline{Fig.1 }:
          Plot of the $F_{C}(v)$ (full line) and $F_{S}(v_{T})$ (dotted line)
	  functions that quantify
          the suppression of the BEC cooling 
          efficiency as compared to cooling by atoms at rest (for $m_{I}/m_{B}
	  = 6/23$). 
	  $F_{C}(v)$ represents 
          the suppression of heat transfer and $F_{S}(v_{T})$
          gives the cooling rate suppression of a Maxwellian
          impurity distribution with thermal velocity $v_{T}$.         
\noindent
\underline{Fig.2 }:
	  Velocity distribution of the impurity atoms as they are cooled
	  by a zero-temperature BEC, plotted at $t=0.4n\tau , 
	  n=0,1,\dots ,10$, $\tau = [4\pi a_{IB}^{2}  n_{B}c
	  (\mu^{2}/m_{I}m_{B})]^{-1}$.  The initial distribution is a 
	  Maxwellian with $v_{T}= 3 c$.  Note that the distribution does
	  not converge to a zero-temperature Maxwellian, but the position of  
	  its maximum converges to a value slightly below $c$: the signature
          of superfluidity.

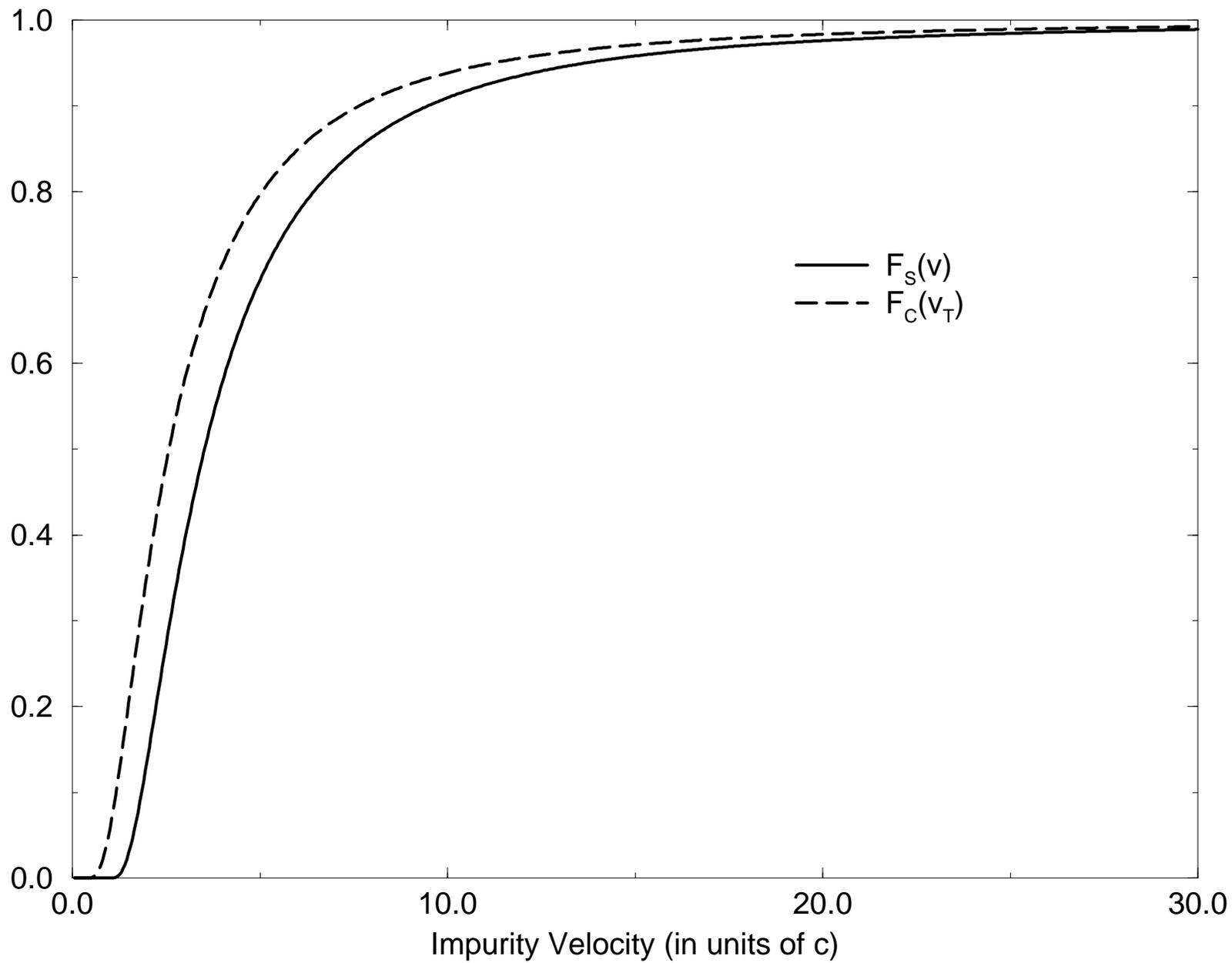
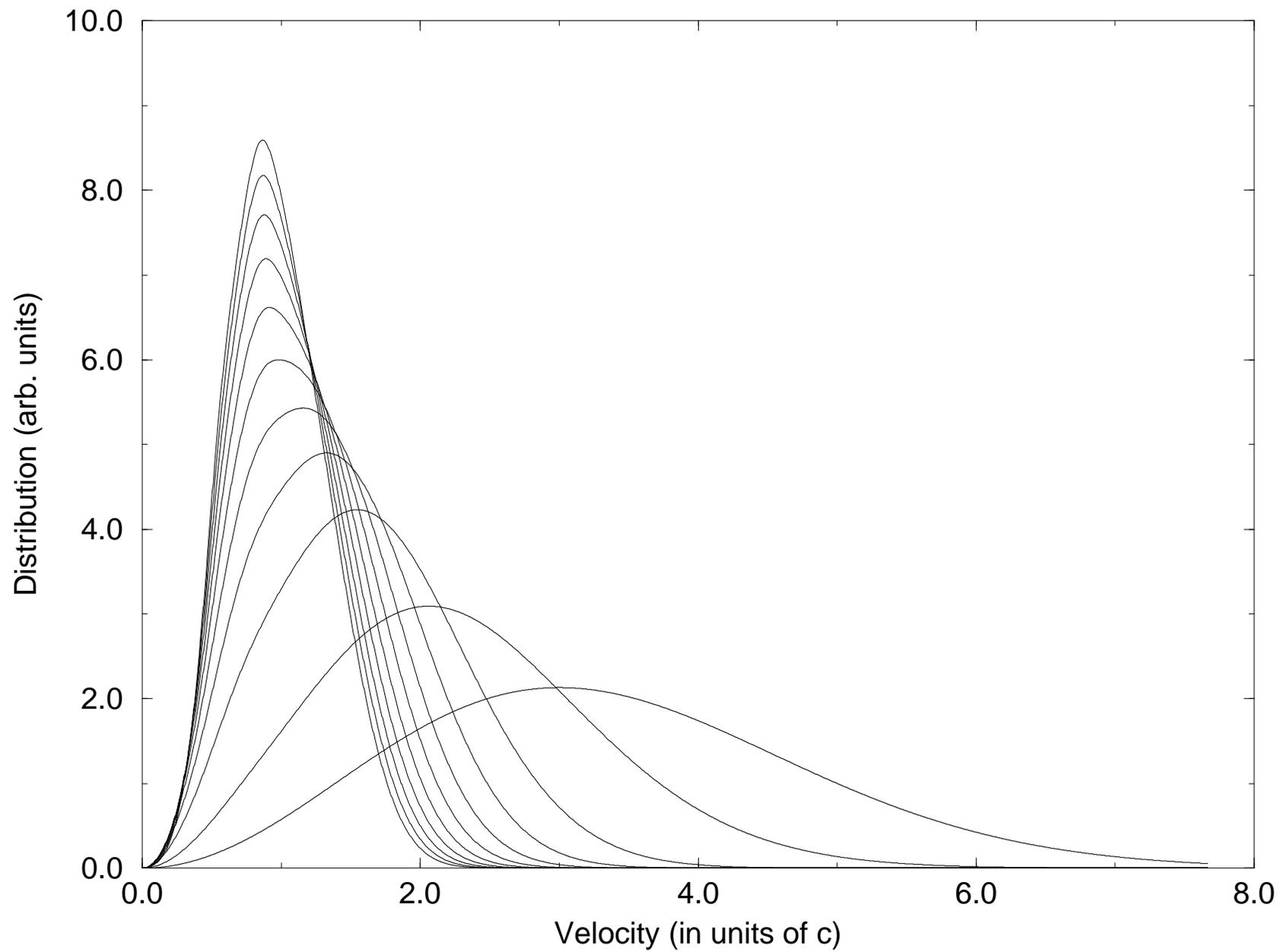
\end{document}